\documentstyle[12pt,psfig]{article}

\begin{document}

\baselineskip=16pt plus 1pt minus 1pt

\begin{center}{\large \bf Ground state bands of the E(5) and X(5) critical 
symmetries obtained from Davidson potentials through a variational procedure } 

\bigskip

{Dennis Bonatsos$^{\#}$\footnote{e-mail: bonat@inp.demokritos.gr},
D. Lenis$^{\#}$\footnote{e-mail: lenis@inp.demokritos.gr}, 
N. Minkov$^\dagger$\footnote{e-mail: nminkov@inrne.bas.bg},
D. Petrellis$^{\#}$,  
P. P. Raychev$^\dagger$\footnote{e-mail: raychev@phys.uni-sofia.bg, 
raychev@inrne.bas.bg}, \break
P. A. Terziev$^\dagger$\footnote{e-mail: terziev@inrne.bas.bg} }
\bigskip

{$^{\#}$ Institute of Nuclear Physics, N.C.S.R.
``Demokritos''}

{GR-15310 Aghia Paraskevi, Attiki, Greece}

{$^\dagger$ Institute for Nuclear Research and Nuclear Energy, Bulgarian
Academy of Sciences }

{72 Tzarigrad Road, BG-1784 Sofia, Bulgaria}

\bigskip

{\bf Abstract}

\end{center} 

Davidson potentials of the form $\beta^2 +\beta_0^4/\beta^2$, when used 
in the original Bohr Hamiltonian for $\gamma$-independent potentials 
bridge the U(5) and O(6) symmetries. Using a variational procedure, 
we determine for each value of angular momentum $L$ the value 
of $\beta_0$ at which the derivative of the energy ratio $R_L =
E(L)/E(2)$ with respect to $\beta_0$ has a sharp maximum, the collection 
of $R_L$ values at these points forming a band which practically coincides 
with the ground state band of the E(5) model, corresponding to the critical 
point in the shape phase transition from U(5) to O(6). 
The same potentials, when used in the Bohr Hamiltonian after separating 
variables as in the X(5) model, bridge the U(5) and SU(3) symmetries,  
the same variational procedure leading to a band which 
practically coincides with the ground state band of the X(5) model, 
corresponding to the critical point of the U(5) to SU(3) shape phase 
transition. A new derivation of the Holmberg--Lipas formula for nuclear energy 
spectra is obtained as a by-product. 

\bigskip

\noindent
PACS: 21.60.Ev; 21.60.Fw 

\newpage 

{\bf 1. Introduction} 

The recently introduced E(5) \cite{IacE5} and X(5) \cite{IacX5} models
are supposed to describe shape phase transitions in atomic nuclei, 
the former being related to the transition from U(5) (vibrational) 
to O(6) ($\gamma$-unstable) nuclei, and the latter corresponding to the 
transition from U(5) to SU(3) (rotational) nuclei. In both cases 
the original Bohr collective Hamiltonian \cite{Bohr} is used, with an infinite 
well potential in the collective $\beta$-variable. 
Separation of variables is achieved in the E(5) case by assuming that the 
potential is independent of the collective $\gamma$-variable, while in the 
X(5) case the potential is assumed to be of the form $u(\beta)+u(\gamma)$.     
We are going to refer to these two cases as ``the E(5) framework'' and 
``the X(5) framework'' respectively.
The selection of an infinite well potential in the $\beta$-variable 
in both cases is justified by the fact that the potential is expected 
to be flat around the point at which a shape phase transition occurs.  
Experimental evidence for the occurence of the E(5) and X(5) symmetries in 
some appropriate nuclei is growing (\cite{CZE5,Zamfir}  and 
\cite{CZX5,Kruecken} respectively). 

In the present work we examine if the choice of the infinite well potential 
is the optimum one for the description of shape phase transitions. 
For this purpose, we need one-parameter potentials which can span 
the U(5)-O(6) region in the E(5) framework, as well as the U(5)-SU(3) region
in the X(5) framework. It turns out that 
the exactly soluble \cite{Elliott,Rowe} Davidson potentials \cite{Dav}  
\begin{equation}\label{eq:e1} 
u(\beta) = \beta^2 +{\beta_0^4 \over \beta^2},
\end{equation}
where $\beta_0$ is the position of the minimum of the potential, 
do possess this property. Taking into account the fact that various 
physical quantities should change most rapidly at the point of the 
shape phase transition \cite{Werner}, we locate for each value of the 
angular momentum $L$ the value of $\beta_0$ for which the rate of change 
of the ratio $R_L=E(L)/E(2)$, a widely used measure of nuclear 
collectivity, is maximized.  It turns out that the collection of $R_L$ 
ratios formed in this way in the case of a potential independent 
of the $\gamma$-variable correspond to the E(5) model, while in the case 
of the $u(\beta)+u(\gamma)$ potential lead to the X(5) model, thus proving 
that the choice of the infinite well potential made in Refs. 
\cite{IacE5,IacX5} is the optimum one. The variational procedure used here 
is analogous to the one used in the framework of the Variable Moment of 
Inertia (VMI) model \cite{VMI}, where the energy is minimized with respect 
to the (angular momentum dependent) moment of inertia for each value 
of the angular momentum $L$ separately. 

In Section 2 the E(5) case is considered, while the X(5) case is examined 
in Section 3, in which a new derivation of the Holmberg--Lipas formula 
\cite{Lipas} for nuclear energy spectra is obtained as a by-product. 
Finally, Section 4 contains a discussion of the present results and 
plans for further work. 

{\bf 2. Davidson potentials in the E(5) framework } 

The original Bohr Hamiltonian \cite{Bohr} is
\begin{equation}\label{eq:e2}
H = -{\hbar^2 \over 2B} \left[ {1\over \beta^4} {\partial \over \partial 
\beta} \beta^4 {\partial \over \partial \beta} + {1\over \beta^2 \sin 
3\gamma} {\partial \over \partial \gamma} \sin 3 \gamma {\partial \over 
\partial \gamma} - {1\over 4 \beta^2} \sum_{k=1,2,3} {Q_k^2 \over \sin^2 
\left(\gamma - {2\over 3} \pi k\right) } \right] +V(\beta,\gamma),
\end{equation}
where $\beta$ and $\gamma$ are the usual collective coordinates describing the 
shape of the nuclear surface,
$Q_k$ ($k=1$, 2, 3) are the components of angular momentum, and $B$ is the 
mass parameter. 

Assuming that the potential depends only on the variable $\beta$, 
i.e. $V(\beta,\gamma) = U(\beta)$, one can proceed to separation of variables 
in the standard way \cite{Bohr,WJ1956}, using the wavefunction 
$ \Psi(\beta,\gamma, \theta_i) = f(\beta) \Phi(\gamma, \theta_i)$,
where $\theta_i$ $(i=1,2,3)$ are the Euler angles describing the orientation 
of the deformed nucleus in space. 

In the equation involving the angles, the eigenvalues of the second order 
Casimir operator of SO(5) occur, having the form 
 $\Lambda = \tau(\tau+3)$, where $\tau=0$, 1, 2, \dots is the quantum 
number characterizing the irreducible representations (irreps) of SO(5), 
called the ``seniority'' \cite{Rakavy}. This equation has been solved 
by Bes \cite{Bes}.   

The ``radial'' equation can be simplified by introducing \cite{IacE5} 
reduced energies $\epsilon = {2B\over \hbar^2} E$ and reduced potentials 
$u= {2B \over \hbar^2} U$, leading to 
\begin{equation} \label{eq:e3} 
\left[ -{1\over \beta^4} {\partial \over \partial \beta} \beta^4 
{\partial \over \partial \beta} + {\tau(\tau+3) \over \beta^2}+ u(\beta) 
\right] f(\beta) = \epsilon f(\beta). 
\end{equation} 

When plugging the Davidson potentials of Eq. (\ref{eq:e1}) in the above 
equation, the $\beta_0^4 /\beta^2$ term is combined with the 
$\tau(\tau+3)/\beta^2$ term appearing there and the equation is solved 
exactly \cite{Elliott,Rowe}, the eigenfunctions 
being Laguerre polynomials of the form 
\begin{equation}\label{eq:e4} 
F^\tau_n(\beta) = \left[{ 2 n! \over \Gamma 
\left( n+p+{5\over 2}\right)}\right]^{1/2} \beta^p L_n^{p+{3\over 2}}(\beta^2)
e^{-\beta^2/2}  
\end{equation}
where $\Gamma(n)$ stands for the $\Gamma$-function, while 
$p$ is determined by \cite{Elliott}
\begin{equation}\label{eq:e5} 
p(p+3) =\tau(\tau+3) + \beta_0^4,
\end{equation} 
leading to 
\begin{equation}\label{eq:e6} 
p= -{3\over 2} + \left[ \left(\tau+{3\over 2}\right)^2 +\beta_0^4\right]^{1/2}.
\end{equation}
The energy eigenvalues are then \cite{Elliott,Rowe}
(in $\hbar \omega=1$ units) 
\begin{equation}\label{eq:e7} 
E_{n,\tau} = 2n+p+{5\over 2} = 
2n+1+ \left[ \left( \tau+{3\over 2} \right)^2 +\beta_0^4
\right]^{1/2} . 
\end{equation}
For $\beta_0=0$ the original solution of Bohr \cite{Bohr}, which corresponds 
to a 5-dimensional (5-D) harmonic oscillator characterized by the symmetry 
U(5) $\supset$ SO(5) $\supset$ SO(3) $\supset$ SO(2) \cite{CM870},  
is obtained. 
The values of angular momentum $L$ contained in each irrep of SO(5) 
(i.e. for each value of $\tau$) are given by the algorithm \cite{IA} 
$\tau=3\nu_\Delta +\lambda$, where $\nu_\Delta=0$, 1, \dots is the missing 
quantum number in the reduction SO(5) $\supset$ SO(3),  and 
$L=\lambda, \lambda+1, \ldots, 2\lambda-2, 2\lambda$ (with $2\lambda-1$ 
missing). 

The levels of the ground state band are characterized by $L=2\tau$ and $n=0$.
Then the energy levels of the ground state band are given by 
\begin{equation}\label{eq:e8} 
E_{0,L} =1+{1\over 2} \left[ (L+3)^2 +4\beta_0^4\right]^{1/2},
\end{equation}
while the excitation energies of the levels of the ground state band 
relative to the ground state are 
\begin{equation}\label{eq:e9} 
E_{0,L,exc}= E_{0,L}-E_{0,0} ={1\over 2} \left( \left[(L+3)^2
+4\beta_0^4\right]^{1/2} -\left[ 9+4\beta_0^4\right]^{1/2} \right).
\end{equation}

For $u(\beta)$ being a 5-D infinite well 
\begin{equation}\label{eq:e10}
  u(\beta) = \left\{ \begin{array}{ll} 0 & \mbox{if $\beta \leq \beta_W$} \\
\infty  & \mbox{for $\beta > \beta_W$} \end{array} \right.
\end{equation} 
one obtains the E(5) model of Iachello \cite{IacE5} in which the 
eigenfunctions are 
Bessel functions $J_{\tau+3/2}(z)$ (with $z=\beta k$, $k =\sqrt{\epsilon}$), 
while the spectrum is determined by the zeros of the Bessel functions 
\begin{equation}\label{eq:e11}  
E_{\xi,\tau} = {\hbar^2 \over 2B} k^2_{\xi,\tau}, \qquad 
k_{\xi,\tau} = {x_{\xi,\tau} \over \beta_W}
\end{equation}
where $ x_{\xi,\tau}$ is the  $\xi$-th zero of the Bessel function 
$J_{\tau+3/2}(z)$. 
The spectra of the E(5) and Davidson cases become directly comparable 
by establishing the formal correspondence $n=\xi-1$. 

It is instructive to consider the ratios 
\begin{equation}\label{eq:e12} 
R_L= {E_{0,L} -E_{0,0} \over E_{0,2}-E_{0,0} },
\end{equation}
where the notation $E_{n,L}$ is used. 

For $\beta_0=0$ it is clear that the original vibrational model 
of Bohr \cite{Bohr} (with $R_4=2$) is obtained, while for large $\beta_0$
the O(6) limit of the Interacting Boson Model (IBM) \cite{IA} 
(with $R_4=2.5$) is approached \cite{Elliott}. The latter fact can be seen 
in Table 1, where the $R_L$ ratios for two different values of the 
parameter $\beta_0$ are shown, together with the O(6) predictions 
(which correspond to $E(L)=A L(L+6)$, with $A$ constant \cite{Casten}). 
It is clear that 
the O(6) limit is approached as $\beta_0$ is increased, the agreement 
being already quite good at $\beta_0=5$. 

It is useful to consider the ratios $R_L$, defined above,
as a function of $\beta_0$. As seen in Fig. 1, where the ratios $R_4$, 
$R_{12}$ and $R_{20}$ are shown, these ratios increase with $\beta_0$, 
the increase 
becoming very steep at some value $\beta_{0,max}$ of $\beta_0$, 
where the first derivative $dR_L\over d\beta_0$ reaches a maximum value, 
while the second derivative $d^2 R_L\over d\beta_0^2$ vanishes. 
Numerical results for $\beta_{0,max}$ are shown in Table 2, together with 
the values of $R_L$ occuring at these points, which are compared to the 
$R_L$ ratios occuring in the ground state band of the E(5) model \cite{IacE5}. 
Very close agreement of the values determined by the procedure described 
above with the E(5) values is observed in Table 2, as well as in Fig. 2,
where these ratios are also shown, together with the corresponding ratios of 
the U(5) and O(6) limits.   

The work performed in this section is reminiscent of a variational procedure. 
Wishing to determine the critical point in the shape phase transition 
from U(5) to O(6), one chooses a potential (the Davidson potential) 
with a free parameter ($\beta_0$), which helps in covering the whole 
range of interest. Indeed, for $\beta_0=0$ the U(5) picture is obtained, while 
large values of $\beta_0$ lead to the O(6) limit.
One then needs a physical quantity which can serve as a ``measure''
of collectivity.  For this purpose one considers 
the ratios $R_L$, encouraged by the fact that these ratios are well-known
indicators of collectivity in nuclear structure \cite{Mallmann}.  
Since at the critical 
point (if any) one expects the collectivity to change very rapidly, one 
looks, for each $R_L$ ratio separately, for the value of the 
parameter at which the change of $R_L$ is maximum. Indeed, the first derivative
of the ratio $R_L$ with respect to the parameter $\beta_0$
exhibits a sharp maximum, which is then a good candidate for being the 
critical point for this particular value of the angular momentum $L$. 
The $R_L$ values at the critical points corresponding 
to each value of $L$ form a collection, which should correspond 
to the behaviour of the ground state band of a nucleus at the critical point.
The infinite well potential used in E(5) succeeds in reproducing all 
these ``critical'' $R_L$ ratios in the ground state band for all 
values of the angular momentum $L$, {\it without using any free parameter}.
It is therefore proved that the infinite well potential is indeed 
the optimum choice for describing the ground state bands of nuclei at the 
critical point of the U(5) to O(6) shape phase transition. 

In other words, starting from the Davidson potentials and using a variational 
procedure, according to which the rate of change of the $R_L$ ratios 
as a function of the parameter $\beta_0$ is 
maximized for each value of the angular momentum $L$ separately, one 
forms the collection of critical values of $R_L$ which corresponds 
to the ground state band of the E(5) model, which is supposed 
to describe nuclei at the critical point.  

Variational procedures in which each value of the angular momentum $L$ is 
treated separately are not unheard of in nuclear physics. An example is 
given by the Variable Moment of Inertia (VMI) model \cite{VMI}, in which 
the energy of the nucleus is minimized with respect to the 
(angular momentum dependent) moment 
of inertia for each value of the angular momentum separately. From 
the cubic equation obtained from this condition, the moment of inertia is 
uniquely determined (as a function of angular momentum) 
in each case. The collection of energy levels occuring 
by using in the energy formula the appropriate value of the 
moment of inertia for each value of the angular momentum $L$ 
forms the ground state band of the nucleus. 

Some comparison of the variational procedure used here with the 
standard Ritz variational method used in quantum mechanics 
(\cite{GM}, for example) is in place. 
In the (simplest version of the) Ritz variational method a trial wave function
containing a parameter is chosen and subsequently the energy is minimized 
with respect to this parameter, thus determining the parameter value 
and, after the relevant substitution, the energy value. 
In the present case a trial potential containing a parameter is chosen 
and subsequently the rate of change of the 
physical quantity (here the rate of change of the energy ratios) 
is maximized with respect to this parameter, thus determining the 
parameter value and, after the relevant calculation, the value of the 
physical quantity (here the energy ratios). 
The main similarity between the two methods is the use of a 
parameter-dependent trial wave function/trial potential respectively. 
The main difference between the two methods is that in the former 
the relevant physical quantity (the energy) is minimized with respect 
to the parameter, while in the latter the rate of change of the physical 
quantity (the energy ratios) is maximized with respect to the parameter.  

{\bf 3. Davidson potentials in the X(5) framework} 

Starting again from the original Bohr Hamiltonian of Eq. (\ref{eq:e2}), 
one seeks solutions of the relevant Schr\"odinger equation having 
the form 
$ \Psi(\beta, \gamma, \theta_i)= \phi_K^L(\beta,\gamma) 
{\cal D}_{M,K}^L(\theta_i)$, 
where $\theta_i$ ($i=1$, 2, 3) are the Euler angles, ${\cal D}(\theta_i)$
denote Wigner functions of them, $L$ are the eigenvalues of angular momentum, 
while $M$ and $K$ are the eigenvalues of the projections of angular 
momentum on the laboratory-fixed $z$-axis and the body-fixed $z'$-axis 
respectively. 

As pointed out in Ref. \cite{IacX5}, in the case in which the potential 
has a minimum around $\gamma =0$ one can write  the last term of Eq. 
(\ref{eq:e2}) in the form 
\begin{equation}\label{eq:e13} 
\sum _{k=1,2,3} {Q_k^2 \over \sin^2 \left( \gamma -{2\pi \over 3} k\right)}
\approx {4\over 3} (Q_1^2+Q_2^2+Q_3^2) +Q_3^2 \left( {1\over \sin^2\gamma}
-{4\over 3}\right).  
\end{equation}
Using this result in the Schr\"odinger equation corresponding to 
the Hamiltonian of Eq. (\ref{eq:e2}), introducing reduced energies 
 $\epsilon = 2B E /\hbar^2$ and reduced potentials $u = 2B V /\hbar^2$,  
and assuming that the reduced potential can be separated into two terms, 
one depending on $\beta$ and the other depending on $\gamma$, i.e. 
$u(\beta, \gamma) = u(\beta) + u(\gamma)$, the Schr\"odinger equation can 
be separated into two equations \cite{IacX5}, the ``radial'' one being   
\begin{equation} \label{eq:e14}
\left[ -{1\over \beta^4} {\partial \over \partial \beta} \beta^4 
{\partial \over \partial \beta} + {1\over 4 \beta^2} {4\over 3} 
L(L+1) +u(\beta) \right] \xi_L(\beta) =\epsilon_\beta  \xi_L(\beta). 
\end{equation}

When plugging the Davidson potentials of Eq. (\ref{eq:e1}) in this equation, 
the $\beta_0^4 /\beta^2$ term of the potential is combined with the 
$L(L+1)/3\beta^2$ term appearing there and the equation is solved exactly, 
the eigenfunctions being Laguerre polynomials of the form 
\begin{equation}\label{eq:e15} 
F^L_n(\beta) = \left[{ 2 n! \over \Gamma 
\left( n+a+{5\over 2}\right)}\right]^{1/2} \beta^a L_n^{a+{3\over 2}}(\beta^2)
e^{-\beta^2/2}  
\end{equation}
where $a$ is given by
\begin{equation}\label{eq:e16} 
a= -{3\over 2} + \left[ {1\over 3} L(L+1) + {9\over 4} +\beta_0^4\right]^{1/2}.
\end{equation}
The energy eigenvalues are then
(in $\hbar \omega=1$ units) 
\begin{equation}\label{eq:e17} 
E_{n,L} = 2n+a +{5\over 2} = 
2n+1+ \left[ {1\over 3} L(L+1) + {9\over 4} +\beta_0^4
\right]^{1/2} . 
\end{equation}

The levels of the ground state band are characterized by $n=0$.
Then the excitation energies relative to the ground state are 
given by 
\begin{equation}\label{eq:e18} 
E_{0,L,exc} = \left[{1\over 3} L(L+1) + {9\over 4} +\beta_0^4\right]^{1/2}
-\left[ {9\over 4} +\beta_0^4\right]^{1/2},
\end{equation}
which can easily be put into the form 
\begin{equation}\label{eq:e19} 
E'_{0,L,exc} = {E_{0,L,exc} \over \left[ {9\over 4} +\beta_0^4\right]^{1/2} }
= \left[1+ {L(L+1)\over 3 \left( {9\over 4} +\beta_0^4 
\right) }\right]^{1/2} -1,
\end{equation}
which is the same as the Holmberg--Lipas formula \cite{Lipas} 
\begin{equation}\label{eq:e20}
E_{H}(L) = a_{H} \left( \sqrt{1+b_{H}  L(L+1)} -1 \right), 
\end{equation}
with $a_H=1$ 
\begin{equation}\label{eq:e21}
b_H = {1\over 3\left( {9\over 4} +\beta_0^4\right)  } . 
\end{equation}

It is clear that the Holmberg--Lipas formula gives rotational spectra 
for small values of $b_H$, at which one can keep only the first 
$L$-dependent term 
in the Taylor expansion of the square root appearing in Eq. (\ref{eq:e20}),
leading to energies proportional to $L(L+1)$. From Eq. (\ref{eq:e21})
it is then clear that rotational spectra are expected for large values 
of $\beta_0$. This can be seen in Table 3, where the $R_L$ ratios 
occuring for two different values of $\beta_0$ are shown, together 
with the predictions of the SU(3) limit of IBM, which correspond 
to the pure rotator with $E(L)=A L(L+1)$, where $A$ constant \cite{IA}. 
The agreement to the SU(3) results is quite good already at $\beta_0=5$. 
On the other hand, 
the case $\beta_0=0$ corresponds to an exactly soluble 
model with $R_4 = 2.646$, which has been called the X(5)-$\beta^2$ 
model \cite{X5}. 

It is worth remarking at this point that the Holmberg--Lipas formula
can be derived \cite{Casten} by assuming that the moment of inertia $I$ in the 
energy expression of the rigid rotator ($E(L)=L(L+1)/2I$) is a function of the 
excitation energy, i.e. $I= \alpha + \beta E(L)$, where $\alpha$ and 
$\beta$ are constants, the latter being proportional to $b_H$ and 
acquiring positive values. It is therefore clear that the Holmberg--Lipas 
formula, as well as the spectrum of the Davidson potentials derived 
in this section, have built-in the concept of the Variable Moment 
of Inertia (VMI) model \cite{VMI}, according to which the moment of inertia 
is an increasing function of the angular momentum. 

For $u(\beta)$ being a 5-D infinite well potential (see Eq. (\ref{eq:e10})
one obtains the X(5) model of Iachello \cite{IacX5}, 
in which the eigenfunctions are Bessel functions $J_\nu(k_{s,L}\beta)$ with 
\begin{equation}\label{eq:e22}
\nu=\left( {L(L+1)\over 3}+{9\over 4}\right)^{1/2},
\end{equation}
while the spectrum is determined by the zeros of the Bessel functions, 
the relevant eigenvalues being
\begin{equation}\label{eq:e23}
\epsilon_{\beta; s,L} = (k_{s,L})^2, \qquad 
k_{s,L}=  {x_{s,L} \over \beta_W},
\end{equation}
where $x_{s,L}$ is the $s$-th zero of the Bessel function 
$J_\nu(k_{s,L}\beta)$. The spectra of the X(5) and Davidson cases become 
directly comparable by establishing the formal correspondence 
$n=s-1$. 

It is useful to consider the ratios $R_L$, defined in the previous 
section, as a function of $\beta_0$. As seen in Fig. 3, these ratios again 
increase with $\beta_0$, the increase becoming very steep at some value 
$\beta_{0,max}$ of $\beta_0$, 
where the first derivative $dR_L\over d\beta_0$ reaches a maximum value, 
while the second derivative $d^2 R_L\over d\beta_0^2$ vanishes. 
Numerical results for $\beta_{0,max}$ are shown in Table 4, together with 
the values of $R_L$ occuring at these points, which are compared to the 
$R_L$ ratios occuring in the ground state band of the X(5) model \cite{IacX5}. 
Very close agreement of the values determined by the procedure described 
above with the X(5) values is observed.   

The work performed here is reminiscent of a variational procedure,
as in the previous section.  
Wishing to determine the critical point in the shape phase transition 
from U(5) to SU(3), one chooses a potential (the Davidson potential) 
with a free parameter ($\beta_0$), which serves in spanning the range 
of interest. For large values of $\beta_0$ the SU(3) limit is obtained, 
while for $\beta_0=0$ the X(5)-$\beta^2$ picture is obtained \cite{X5}, 
which is not the U(5) limit, but it is located between U(5) and X(5), 
on the way from U(5) to SU(3). Thus the region of interest around X(5) 
is covered from X(5)-$\beta^2$ to SU(3).  
Then the values of $\beta_0$ at which the first derivative $dR_L/d\beta_0$
exhibits a sharp maximum are determined for each value of the angular 
momentum $L$ separately, the collection of $R_L$ ratios at these values of 
$\beta_0$ forming a band, which turns out to be in very good agreement 
with the ground state band of X(5), the model supposed to be appropriate 
for describing nuclei at the critical point in the transition from 
U(5) to SU(3), thus indicating that the choice of the infinite well 
potential used in the X(5) model is the optimum one. 
The results are depicted in Fig. 4, where in addition to the bands 
provided by the variational procedure and the X(5) model, the bands 
corrsponding to the U(5), X(5)-$\beta^2$, and SU(3) cases are shown.

{\bf 4. Discussion}

The main results and conclusions obtained in  the present work 
are listed here: 

1) A variational procedure for determining the values of physical 
quantities at the point of shape phase transitions in nuclei 
has been suggested. Using one-parameter potentials spanning the region 
between the two limiting symmetries of interest, the parameter values 
at which the rate of change of the physical quantity becomes maximum are 
determined for each value of the angular momentum separately and the 
corresponding values of the physical quantity at these parameter 
values are calculated. The values of the physical quantity collected in this 
way represent its behaviour at the critical point.

2) The method has been applied in the shape phase transition from U(5) to O(6),
using one-parameter Davidson potentials \cite{Dav} and considering 
the energy ratios $R_L=E(L)/E(2)$ within the ground state band as the 
relevant physical quantity, leading to a band which practically coincides 
with the ground state band of the E(5) model \cite{IacE5}. 
It has also been applied in the same way in the shape phase transition from 
U(5) to SU(3), leading to a band which practically coincides with 
the ground state band of the X(5) model \cite{IacX5}. 

3) It should be 
emphasized that the application of the method was possible because 
the Davidson potentials correctly reproduce the U(5) and O(6) symmetries 
in the former case (for small and large parameter values respectively), 
as well as the relevant X(5)-$\beta^2$ \cite{X5} and SU(3) symmetries 
in the latter case (for small and large parameter values respectively).     

4) As a by-product, a derivation of the Holmberg--Lipas formula \cite{Lipas} 
has been achieved using Davidson potentials in the X(5) framework. 

It is clearly of interest to apply the variational procedure introduced here 
to physical quantities other than the energy ratios in the ground state band. 
Energy ratios involving levels of excited bands, ratios of B(E2) transition 
rates (both intraband and interband), and ratios of quadrupole moments
are obvious choices. Work in these directions is in progress, using 
the Davidson potentials, since they possess the appropriate limiting 
behaviour for small and large parameter values. However, any other 
potential/Hamiltonian bridging the relevant pairs of symmetries 
(U(5)-O(6) and U(5)-SU(3)) should be equally appropriate. 
 
{\bf Acknowledgements}

Partial support through the NATO Collaborative Linkage Grant PST.CLG 978799 is 
gratefully acknowledged.

\bigskip 

\centerline{\bf Figure captions} 

{\bf Fig. 1} The $R_L$ ratios (defined in Eq. (\ref{eq:e12})) 
for $L=4$, 12, 20 and their derivatives $dR_L/d\beta_0$ vs. the parameter 
$\beta_0$,
calculated using Davidson potentials (Eq. (\ref{eq:e1})) in the E(5) framework.
The $R_L$ curves also demonstrate the evolution from the U(5) symmetry (on the
left) to the O(6) limit (on the right). See section 2 for further details. 

{\bf Fig. 2}  Values of the ratio $R_L$ (defined in Eq. (\ref{eq:e12}))
obtained through the variational procedure (labeled by ``var'') using 
Davidson potentials in the E(5) framework, compared to the values provided 
by the U(5), O(6), and E(5) models. See section 2 for further details. 

{\bf Fig. 3} The $R_L$ ratios (defined in Eq. (\ref{eq:e12})) 
for $L=4$, 12, 20 and their derivatives $dR_L/d\beta_0$ vs. the parameter 
$\beta_0$, calculated using Davidson potentials (Eq. (\ref{eq:e1})) in the 
X(5) framework. The $R_L$ curves also demonstrate the evolution from the 
X(5)-$\beta^2$ symmetry (on the left) to the SU(3) limit (on the right). 
See section 3 for further details. 

{\bf Fig. 4} Values of the ratio $R_L$ (defined in Eq. (\ref{eq:e12}))
obtained through the variational procedure (labeled by ``var'') using 
Davidson potentials in the X(5) framework, compared to the values provided 
by the U(5), SU(3), X(5), and X(5)-$\beta^2$  models. See section 3 
for further details. 

\newpage 
\parindent=0pt

\begin{table}

\caption{$R_L$ ratios (defined in Eq. (\ref{eq:e12}))
for the ground state band of the Davidson potentials in the E(5) framework 
(Eq. (\ref{eq:e8})) for different values of the parameter 
$\beta_0$, compared to the O(6) exact results.  
}

\bigskip

\begin{tabular}{r r r r}
\hline
L  &  $R_L$       &  $R_L$        & $R_L$ \\
   & $\beta_0=5.$ & $\beta_0=10.$ & O(6) \\
\hline
4  &  2.494 &  2.500 &  2.500 \\
6  &  4.475 &  4.498 &  4.500 \\
8  &  6.935 &  6.996 &  7.000 \\
10 &  9.861 &  9.991 & 10.000 \\
12 & 13.242 & 13.483 & 13.500 \\
14 & 17.064 & 17.471 & 17.500 \\
16 & 21.312 & 21.954 & 22.000 \\
18 & 25.969 & 26.930 & 27.000 \\
20 & 31.020 & 32.398 & 32.500 \\
\hline 
\end{tabular}
\end{table}

\newpage 
\parindent=0pt

\begin{table}

\caption{Parameter values $\beta_{0,max}$ where the first derivative 
of the energy ratios $R_L$ (defined in Eq. (\ref{eq:e12})) in the E(5) 
framework has a maximum, while the second derivative vanishes,
together with the $R_L$ ratios obtained at these values (labeled by ``var'')
and the corresponding ratios of the E(5) model, for several 
values of the angular momentum $L$. 
}

\bigskip

\begin{tabular}{r r r r}
\hline
L & $\beta_{0,max}$ & $R_L$ & $R_L$ \\
  &               & var   &  E(5) \\
\hline
4  & 1.421 &  2.185 &  2.199 \\
6  & 1.522 &  3.549 &  3.590 \\
8  & 1.609 &  5.086 &  5.169 \\
10 & 1.687 &  6.793 &  6.934 \\
12 & 1.759 &  8.667 &  8.881 \\
14 & 1.825 & 10.705 & 11.009 \\
16 & 1.888 & 12.906 & 13.316 \\
18 & 1.947 & 15.269 & 15.799 \\
20 & 2.004 & 17.793 & 18.459 \\
\hline 
\end{tabular}
\end{table}

\newpage 
\parindent=0pt

\begin{table}

\caption{$R_L$ ratios (defined in Eq. (\ref{eq:e12}))
for the ground state band of the Davidson potentials in the X(5) framework 
(Eq. (\ref{eq:e17})) for different values of the parameter 
$\beta_0$, compared to the SU(3) exact results.  
}

\bigskip

\begin{tabular}{r r r r}
\hline
L  &  $R_L$       &  $R_L$        & $R_L$ \\
   & $\beta_0=5.$ & $\beta_0=10.$ & SU(3) \\
\hline
4  &  3.327 &  3.333 &  3.333 \\
6  &  6.967 &  6.998 &  7.000 \\
8  & 11.897 & 11.993 & 12.000 \\
10 & 18.087 & 18.317 & 18.333 \\
12 & 25.503 & 25.968 & 26.000 \\
14 & 34.102 & 34.941 & 35.000 \\
16 & 43.839 & 45.233 & 45.333 \\
18 & 54.665 & 56.841 & 57.000 \\
20 & 66.530 & 69.760 & 70.000 \\
\hline 
\end{tabular}
\end{table}

\newpage 
\parindent=0pt

\begin{table}

\caption{Parameter values $\beta_{0,max}$ where the first derivative 
of the energy ratios $R_L$ (defined in Eq. (\ref{eq:e12})) in the X(5) 
framework has a maximum, while the second derivative vanishes,
together with the $R_L$ ratios obtained at these values (labeled by ``var'')
and the corresponding ratios of the X(5) model,  
for several values of the angular momentum $L$. 
}

\bigskip

\begin{tabular}{r r r r}
\hline
L & $\beta_{0,max}$ & $R_L$ & $R_L$ \\
  &               & var   &  X(5) \\
 \hline
4  & 1.334 &  2.901 &  2.904 \\
6  & 1.445 &  5.419 &  5.430 \\
8  & 1.543 &  8.454 &  8.483 \\
10 & 1.631 & 11.964 & 12.027 \\
12 & 1.711 & 15.926 & 16.041 \\
14 & 1.785 & 20.330 & 20.514 \\
16 & 1.855 & 25.170 & 25.437 \\
18 & 1.922 & 30.442 & 30.804 \\
20 & 1.985 & 36.146 & 36.611 \\
\hline 
\end{tabular}
\end{table}

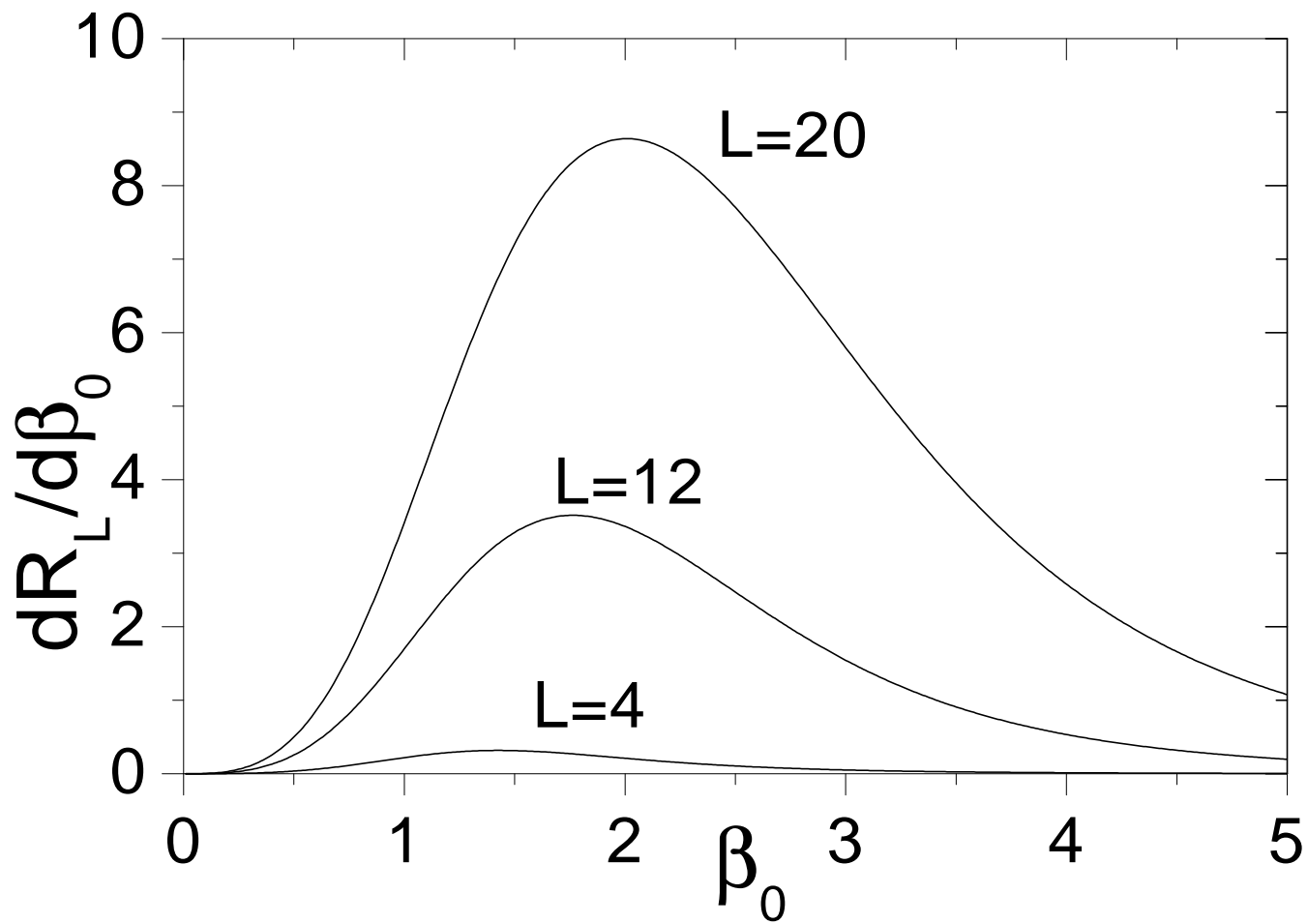
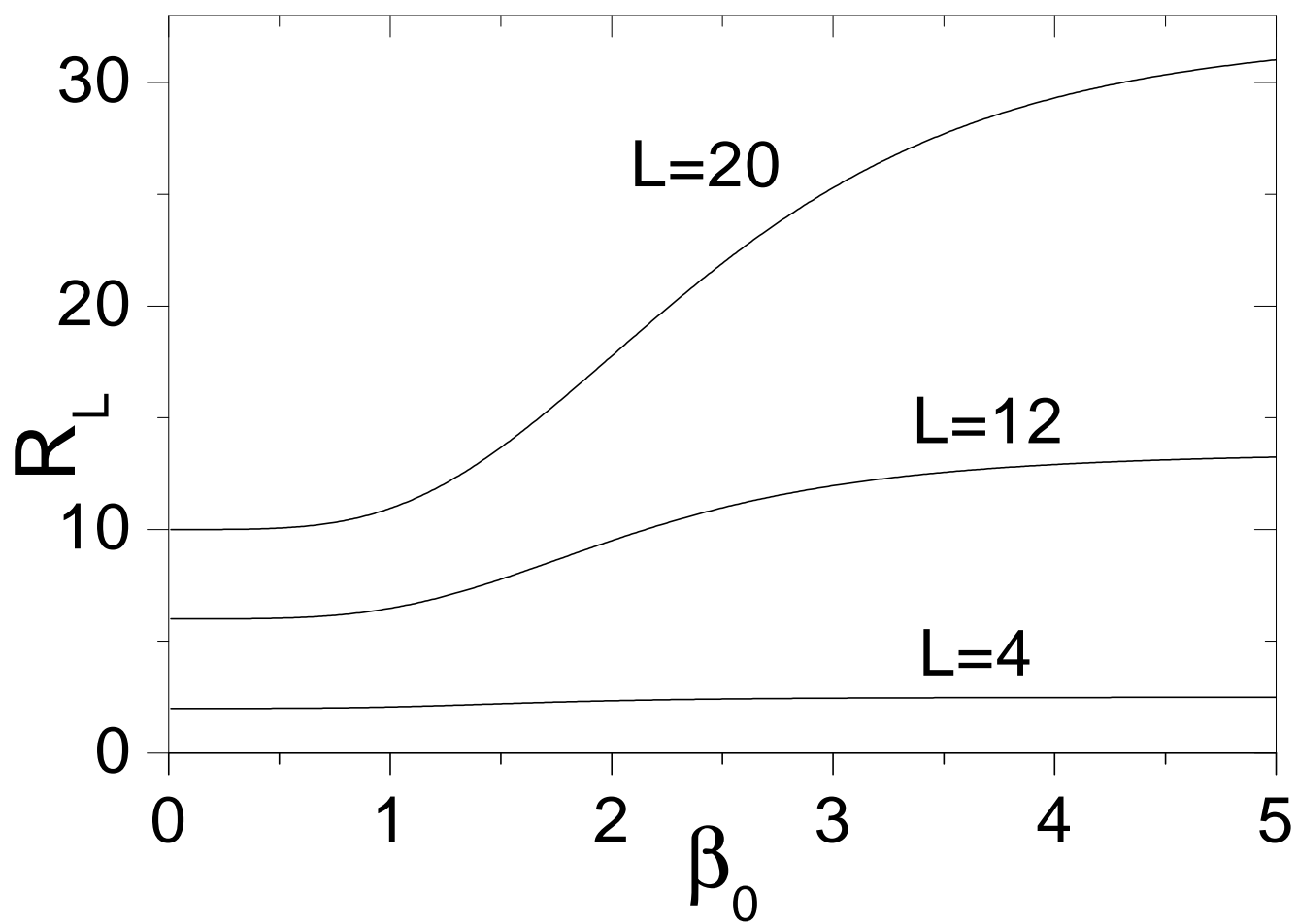
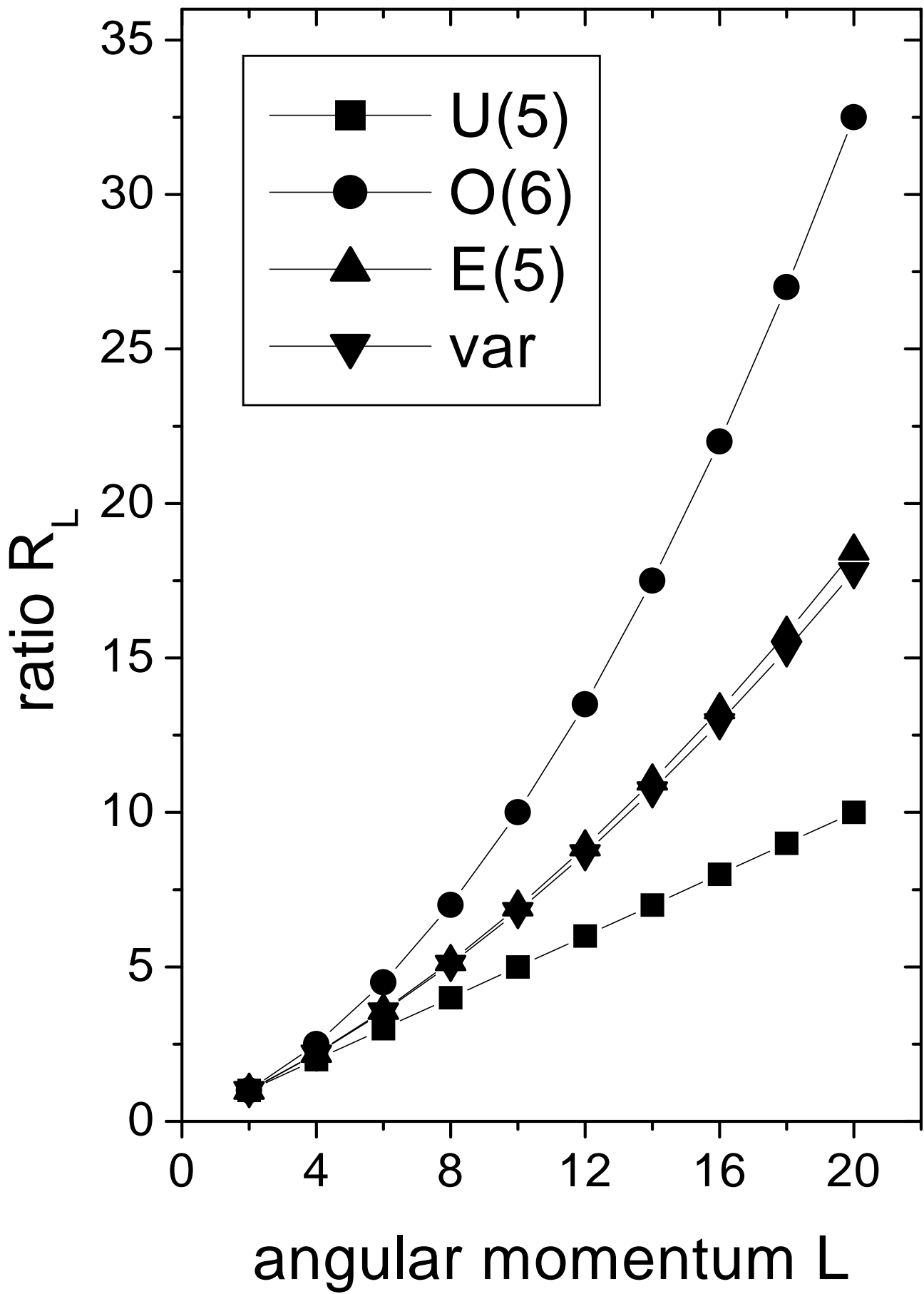
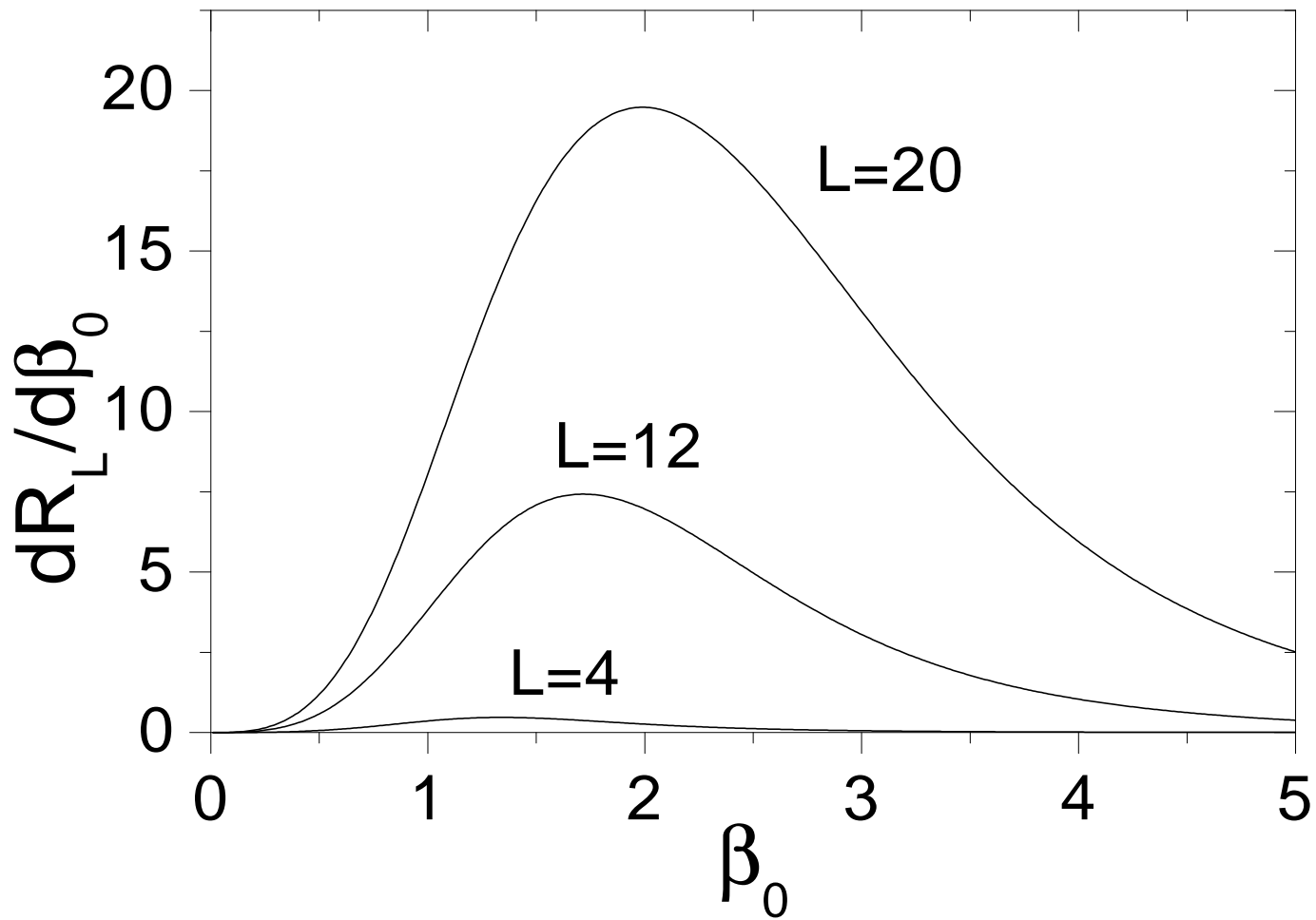
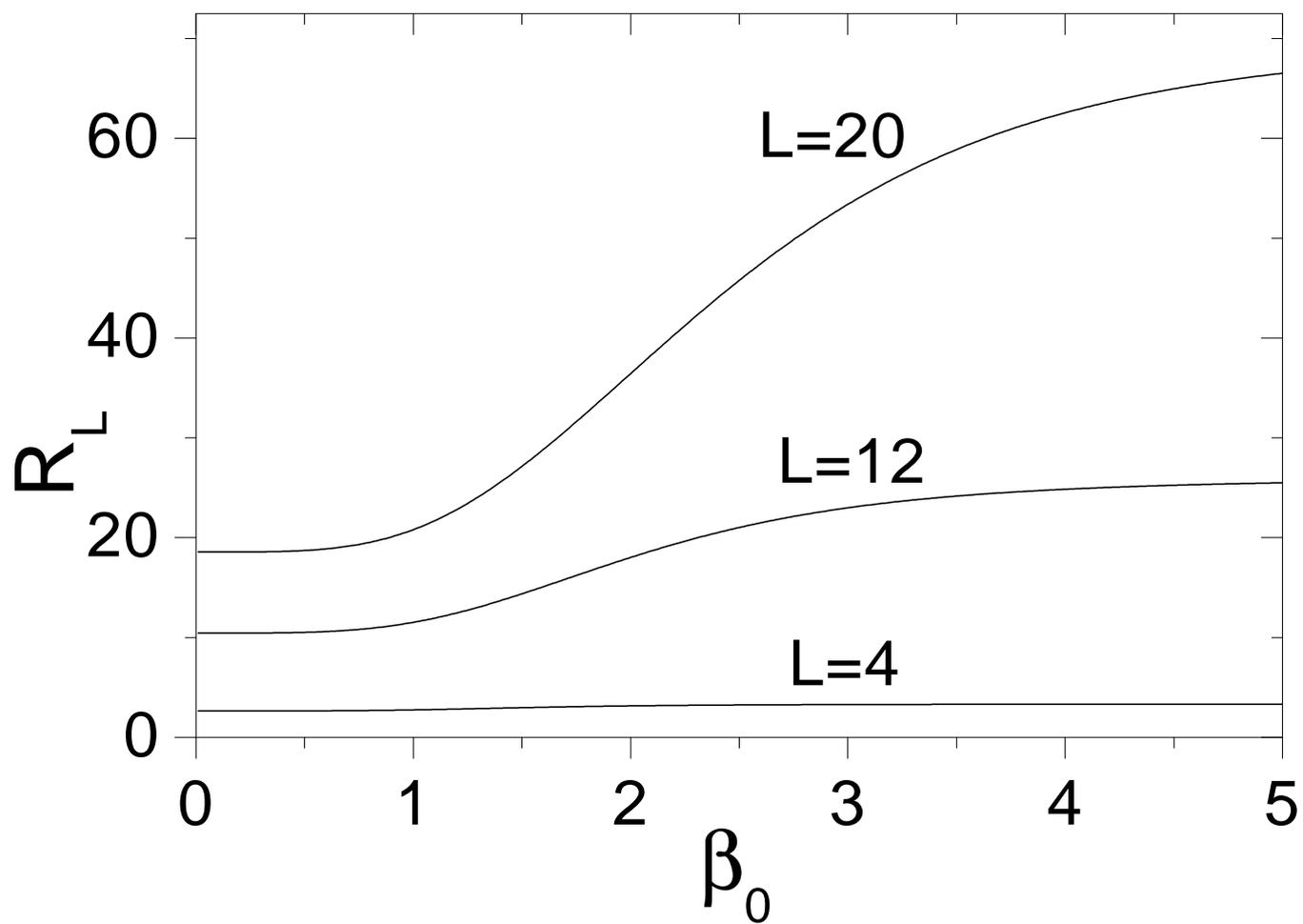
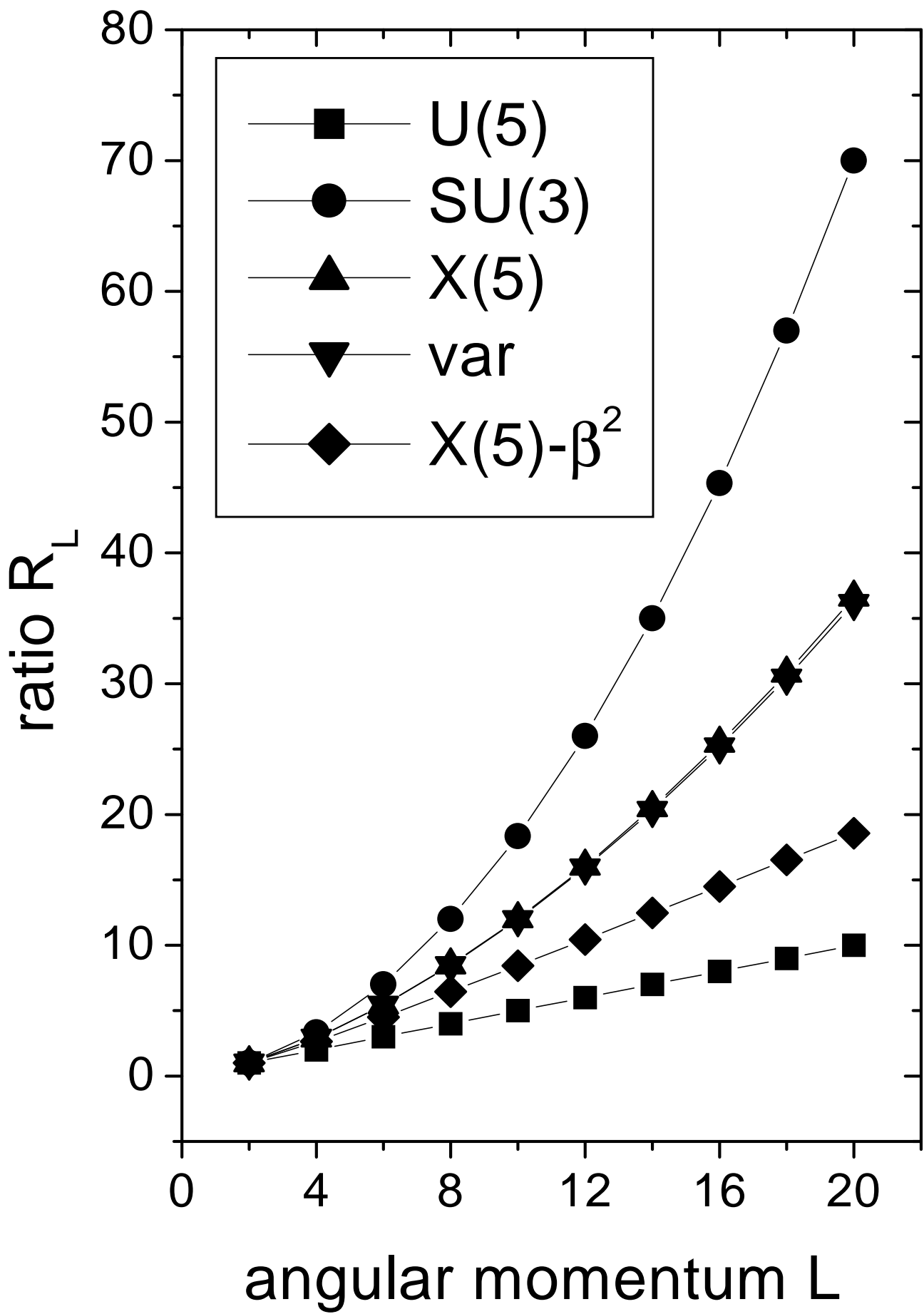
\end{document}